\documentclass[twocolumn,showpacs,prb]{revtex4}
\newcommand{\figurewidth}{\columnwidth}
\usepackage{graphicx}

\begin{document}

\title{Energetics of clusters in the two-dimensional Ising spin glass}

\author{Ludovic Berthier}
\email{berthier@thphys.ox.ac.uk}
\homepage{http://www-thphys.physics.ox.ac.uk/users/LudovicBerthier}
\affiliation{Theoretical Physics,
1 Keble Road, Oxford OX1 3NP, UK and \\
Laboratoire des Verres, Universit\'e Montpellier II, 34095
Montpellier, France}

\author{A.~P.~Young}
\email{peter@bartok.ucsc.edu}
\homepage{http://bartok.ucsc.edu/peter}
\affiliation{Department of Physics,
University of California,
Santa Cruz, California 95064}

\date{\today}

\begin{abstract}
We study numerically the properties of
local low-energy excitations in the two-dimensional 
Ising spin glass. Given the ground state, we
determine the lowest-lying connected cluster of flipped spins 
containing one given spin, either with a fixed 
volume, or with a volume constrained to lie in a certain range.
Our aim is to understand corrections to 
the scaling predicted by the droplet picture of spin glasses and
to resolve contradictory results reported in the literature 
for the stiffness exponent. We find no clear trace of 
corrections to scaling, and the obtained stiffness exponent  
is in relatively good agreement with standard domain wall calculations. 
\end{abstract}

\maketitle

\section{Introduction}
\label{introduction}

According to one of the principal scenarios for the spin glass state,
the ``droplet picture''\cite{fisher:86,fisher:87,fisher:88,bray:86},
the minimum energy excitation of linear dimension $L$
containing a given spin, a droplet, has an energy proportional to
$L^{\theta}$. The ``stiffness exponent'' $\theta$ is positive if
one assumes that the transition
temperature $T_c$ is finite. 
It follows that, in the thermodynamic limit,
excitations that flip a finite fraction of the spins cost an infinite
energy.
However, the results of
several numerical
calculations\cite{krzakala:00,palassini:00,marinari:00,katzgraber:01,katzgraber:01a,katzgraber:01c,houdayer:00,houdayer:00b},
on small system sizes, imply that
the amount of energy needed to
generate system-size droplet excitations is independent of size,
so that $\theta$
for droplet-like excitations is zero. This is
consistent with the alternative replica symmetry breaking 
(RSB)\cite{parisi:79,parisi:80,parisi:83,mezard:87} picture.
Note that numerics simultaneously finds that the 
stiffness exponent for domain-wall excitations is positive and
therefore apparently different from the exponent for droplets, 
an unexpected feature.

To explain the discrepancy between those two
different estimates of the stiffness exponent,  
it has been proposed\cite{moore:02,hartmann:02b}
that the droplet theory is correct on
large scales, i.e.~the stiffness exponent $\theta$ is the same for droplets
and domain walls, and that the apparently contradictory
numerical data can be explained by corrections to
scaling. More precisely Refs.~\onlinecite{moore:02,hartmann:02b} propose that
the droplet energy $\Delta E$ has the form
\begin{equation}
\Delta E  = A L^\theta + B L^{-\omega} ,
\label{deltaE}
\end{equation}
where $\theta > 0$ (if $T_c > 0$ which is the case for $d \ge 3$), $\omega$
is a correction to scaling exponent, and $A$ and $B$ are positive constants. 
If $\theta > 0$, $\Delta E$ has a minimum at some value of
$L$ and if this occurs in the
range of sizes
where numerical data is taken, the numerical values could be roughly
independent of size in the range studied.  In general, only a modest range of
sizes, $L \lesssim 10$, can be studied in dimension $d$ equal to 3.

To quantitatively check this proposal,
simulations have recently
been carried out\cite{katzgraber:03} on a
one-dimensional model with long range interactions which fall off as a power
of the distance, in a region of parameters for which $T_c > 0$.
A much larger range of sizes can be studied than in three
dimensions but even up to $L=512$ the energy of system-size excitations
was found to be independent of size. Hence, for Eq.~(\ref{deltaE}) to be
correct with $\theta > 0$
the values of $\theta$, $\omega$, $A$ and $B$ must conspire to give an
almost constant $\Delta E$ over a very wide range of sizes.

Another model where large sizes can be studied is the two-dimensional
Ising spin
glass at $T=0$, where efficient
algorithms\cite{bieche:80,hartmann:01,rieger:96}
can be applied to determine exact ground
states.  Now $\theta < 0$ in $d=2$, with
$T=0$ ``domain-wall'' calculations
giving\cite{mcmillan:84b,bray:84,rieger:96,palassini:99a,hartmann:01a,carter:02}
$\theta \simeq -0.28$.
However, finite-$T$ simulations\cite{liang:92,kawashima:92a}
and $T=0$ studies
in a magnetic field\cite{kawashima:92},
both of which excite droplets rather
than domain walls, find $\theta \simeq -0.47$. A good review of this
situation is given in Ref.~\onlinecite{kawashima:00a}. Since $\theta <0$ in
$d=2$, both the $L^\theta$ and $L^{-\omega}$
contributions to the droplet energy in Eq.~(\ref{deltaE})
decrease with increasing $L$,
but Refs.~\onlinecite{moore:02,hartmann:02b} argue that they
could combine to give an effective exponent of about $-0.47$ for
small $L$ crossing over to the asymptotic value of about $-0.28$ for larger
sizes. Hartmann and Moore\cite{hartmann:02b} find evidence for this crossover
in their numerical studies in which a certain prescription was used to generate
droplets, though earlier
Hartmann and one of us\cite{hartmann:02}, who generated droplets
in a different way, did \textit{not}
find evidence for an effective exponent close
to $-0.47$. This difference
may indicate that the amplitude of the correction term $B$ in
Eq.~(\ref{deltaE})
depends significantly on the precise way in which
the droplets are generated.
In yet another way of generating droplets, Picco
et al.\cite{picco:02,picco:03} find $\theta$ close to the $-0.47$ value but
with \textit{no} crossover to the domain wall value of around $-0.28$
for the larger sizes. 

Recently, in an interesting paper, Lamarcq et al.\cite{lamarcq:02} have
calculated the energy of droplet (cluster)\cite{cluster_defn}
excitations in three dimensions, in a manner
very similar to the spirit of the droplet
model\cite{fisher:86,fisher:87,fisher:88}. Interestingly they find that
the energy actually \textit{decreases}, 
although very slightly, with increasing size. They
also find that the volume of the droplets has a non trivial fractal dimension
less than the space dimension.
In Ref.~\onlinecite{moore:02}, this slow decrease of the droplet energy 
was taken as a possible evidence of the relevance of 
the correction to scaling term in Eq.~(\ref{deltaE}). Note
however that no sign of a crossover towards the supposedly correct value
of the stiffness exponent was reported, even 
when larger clusters could later be included 
in the analysis\cite{lamarcq:03}.

Here we perform a calculation 
similar to that of Lamarcq et al. but in two
dimensions. The aim is to see if the energies of the droplet energies
calculated in this way fit Eq.~(\ref{deltaE}) and give an effective exponent
close to $-0.47$ at small sizes crossing over to about $-0.28$ at larger sizes
as has been proposed by Moore\cite{moore:02} and Hartmann and
Moore{\cite{hartmann:02b}.
Since the way our clusters are generated is 
directly inspired by the original definition of the 
droplets, one could expect the prediction (\ref{deltaE}) 
to be well suited in our case. Instead, we do not find
any crossover in $\Delta E$ 
by applying the definition of Lamarcq et al. in 2$d$ 
for clusters as large as 64 spins, neither do we find a very negative
exponent $-0.47$. 
We find however an exponent similar to $-0.47$ at very small sizes
if a slight 
modification of the definition of the clusters is made, 
namely if the size of the clusters is let free to evolve between
$n$ and $2n$, as suggested by the 
original ``scale invariant'' definition of the droplets.

\section{The Model and Details of the Numerics}
\label{model}
We take the standard Edwards-Anderson spin glass model
\begin{equation}
{\cal H} = -\sum_{\langle i, j \rangle} J_{ij} S_i S_j,
\label{ham}
\end{equation}
where the $S_i = \pm 1$ are Ising spin variables
at the sites of a simple cubic lattice, and the $J_{ij}$ are nearest neighbor
interactions with a Gaussian distribution with
zero mean and standard deviation unity.  Periodic boundary
conditions are applied on lattices with $N=L^2$ spins. For most of our work we
take $L=64$ but we also did some calculations with smaller sizes down 
to $L=16$. We use $N_s = 1000$
realizations of the disorder.

We are interested in the properties of
low-energy, droplet-like excitations above the ground state
of the Hamiltonian in Eq.~(\ref{ham}). In order to determine these,
we first find the ground state
for each realization of the disorder.
This is done using either
the K\"oln spin glass server\cite{spin_glass_server}, or, for small sizes,
by parallel tempering~\cite{hukushima:96,marinari:98b}.

Following Ref.~\onlinecite{lamarcq:02},
we then generate a droplet by first
choosing randomly a ``central'' spin in the system and reversing it.
We then construct a connected cluster of fixed size $n$
around this central spin by flipping all the spins in this cluster.
We then find the new ground state with the following three constraints:
\begin{enumerate}
\item The central spin is always flipped with respect to the ground state,
\item The number of spins in the cluster is always $n$,
\item The cluster is always connected.
\end{enumerate}
The new ground state is found by a combination of
parallel tempering
and a Kawasaki-type dynamics for the spins in the cluster, in order 
to conserve its size $n$ constant.
Our algorithm does not satisfy detailed balance, but this is
irrelevant since we only want to find the new, constrained ground state.
The lowest temperature in the parallel tempering
is taken to be sufficiently low, $T=0.02$, that
the system is always in a local minimum with no random noise kicking the
system to higher energy states. Our results for ground states in the
presence of a droplet are obtained from spin
configurations at this lowest
temperature. We study cluster sizes up to $n=64$.

In this procedure, which follows Ref.~\onlinecite{lamarcq:02}, the size of the
cluster is strictly fixed. It is therefore slightly different from the
droplets
of Fisher and Huse\cite{fisher:86,fisher:87,fisher:88} which are defined in a
``scale-invariant'' way, i.e. the size is not strictly fixed but allowed
to vary over a certain scale. We have therefore also computed the energies of
scale-invariant excitations in which
the size of the cluster is allowed to vary by a factor of 2, more
precisely to lie between $n/2$ and $n - 1$
for $n= 4$, $8$, $\cdots$, $64$.
Examples of some randomly chosen clusters of size $n=64$
are shown in Fig.~\ref{gif}.

\begin{figure}
\includegraphics[width=8cm,height=8cm]{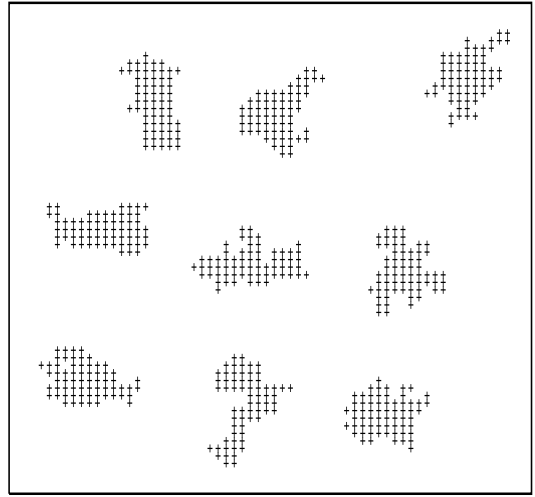}
\caption{Randomly chosen clusters of size $n=64$
obtained by the procedure described in Sec.~\ref{model}.}
\label{gif}
\end{figure}

Our criterion for ensuring that we have
found the true ground state in the presence
of an overturned cluster is as follows.
We first make a few (typically 10) extremely long runs with different bond
configurations to estimate
the typical time scale to reach the new ground state,
$t_{\rm typ}(n)$.
We record the energy versus time of the lowest temperature
for the
two copies, for simulations of at least $10 t_{\rm typ}$ sweeps.
From these graphs,
we can evaluate $t_{\rm typ}$ as the typical time it takes those 10
realizations of the disorder to give a result that can be confidently
taken as the ground state.
Then, for each of the $N_s$ samples, we first run two copies
of the system with independently drawn initial clusters 
for $t_{\rm typ}(n)$ Monte Carlo steps.
We then require 
that the two copies have the same energy for a further continuous period of 
time equal to $t_{\rm typ}(n)$. In this way, we spend more 
time on the ``hard'' samples, which
need several times $t_{\rm typ}(n)$ to converge, than on the easy samples for
which we do not need to run much more than
$t_{\rm typ}(n)$ sweeps.
Obviously, one can not be absolutely sure that the ground state
has really been found since it might happen that both copies
stay for a time larger than $t_{\rm typ}(n)$ in the same
state which is not the
ground state. However, this seems to be fairly unlikely. 
To be on the safe side we
used $t_{\rm typ}(n=64)$, for \textit{all} sizes,
even though, for $n < 64$, one has
$t_{\rm typ}(n) \ll t_{\rm typ}(n=64)$.

For the scale-invariant clusters we start one copy with the minimum-size
cluster, $n/2$, and the other with the maximum-size cluster, $n-1$, to minimize
further the probability that the two copies inadvertently spend a long 
time in the same state which is not the ground state.

We now present our numerical results.

\section{results}
\label{results}

\subsection{Clusters of Fixed Size}
\label{sec:fixed_vol}
In this section we present results of the simulations where we fix the
droplet size precisely, as was done by Lamarcq et al.\cite{lamarcq:02}
on the $3d$ model.

\subsubsection{Droplet Energies}

\begin{figure}
\includegraphics[width=\figurewidth]{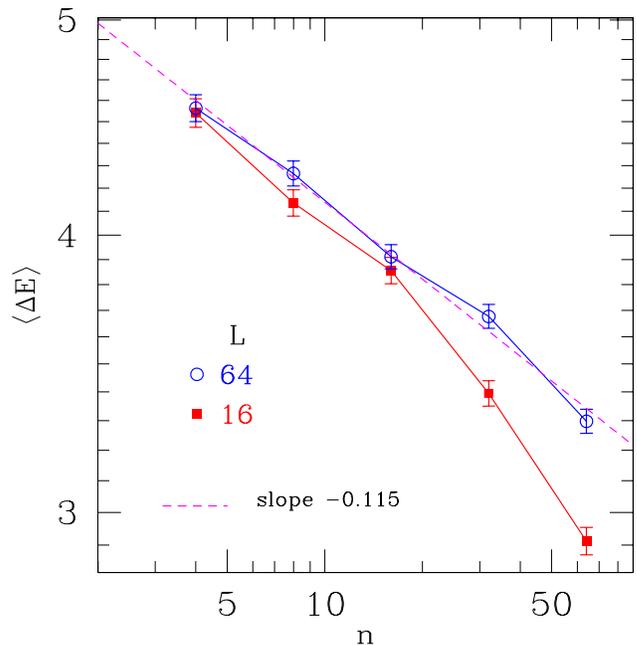}
\caption{
Mean droplet energy against
the number of spins in the droplet for lattice sizes
$L=16$ and 64 for droplets of fixed size.
Data for $L=32$ is very similar to that for $L=64$ and is not
shown for clarity. The slope for $L=64$, which is equal
to $\theta/d_f$, is $-0.115$. No crossover 
in the slope can be detected in the $L=64$ data.
}
\label{fixed_vol}
\end{figure}

Fig.~\ref{fixed_vol} shows the droplet energy, $\langle \Delta E
\rangle$, where the average is over $N_s=1000$
samples, as a function of $n$, the
number of spins in the droplet, on a log-log scale. 
For each sample, $\Delta E$ is the energy difference between the constrained
and unconstrained systems, by definition a positive quantity. 
If we define the fractal
dimension of the volume of the droplets to be $d_f$, i.e.
\begin{equation}
n \sim R^{d_f},
\label{d_f}
\end{equation}
where $R$ is the mean ``radius'' of the droplet (to be defined 
below) then the slope of the data is
$\theta/d_f$, since by definition $\langle \Delta E \rangle
\sim R^\theta$.
 The expectation of the
droplet theory is that $d_f$ is equal to the space dimension, 
$d_f = d = 2$. The slope of
$-0.115$ for $L=64$ would lead to $\theta = -0.23$, not quite as negative 
as the
established value of $-0.28$ from domain wall calculation, but perhaps the
difference is not significant given the rather small range of droplet sizes 
than
we can study. Actually, fits to data presented below indicate that $d_f$ is
somewhat less than two, which would accentuate the difference between 
our value of $\theta$ and $-0.28$. However, the apparent difference
between $d_f$ and 2 may itself be due to corrections to scaling. 

We note also in Fig.~\ref{fixed_vol} that the data for $L=16$ shows an 
interesting feature. For small cluster sizes, the relation between 
$\langle \Delta E \rangle$ and $n$ follows that of the $L=64$ and 
the two sets of data start to depart from each other at large cluster
sizes, so that the apparent stiffness exponent becomes more negative
at large sizes. This crossover is however exactly opposite
to the one expected from Eq.~(\ref{deltaE}). 
A visual inspection reveals that even the clusters with $n=64$ have a
radius smaller than the system size $L=16$, so that the effect
is not trivially due to the fact that boundaries of the clusters interact
with each other. Rather, this crossover
is most probably due a finite size effect on 
the initial ground state energy, a smaller size making it too large, resulting
then in too small an energy difference $\Delta E$.

An important conclusion drawn from Fig.~\ref{fixed_vol} is that
no crossover from a large to a smaller value of the stiffness exponent
is visible, although our data cover a reasonably large size window.
We are led to the conclusion that the definition of the droplets
used here is relatively free of the corrections to scaling expected 
from Eq.~(\ref{deltaE}), and moreover that it leads, for $d=2$, to a 
determination of the stiffness exponent in fair agreement
with domain wall calculations.

\subsubsection{Distribution of Droplet Energies}
\begin{figure}
\includegraphics[width=\figurewidth]{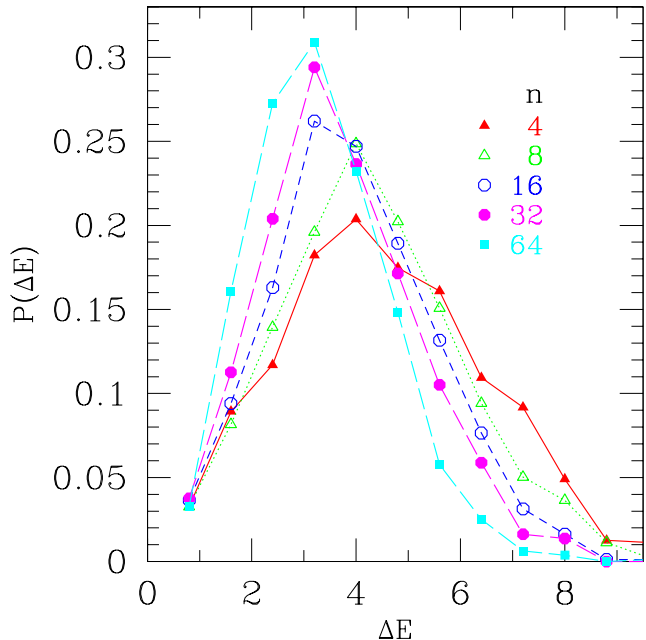}
\caption{Distribution of the energy of the droplets of fixed size $n$, 
for different $n$ and system size $L=64$.}
\label{dist}
\end{figure}

\begin{figure}
\includegraphics[width=\figurewidth]{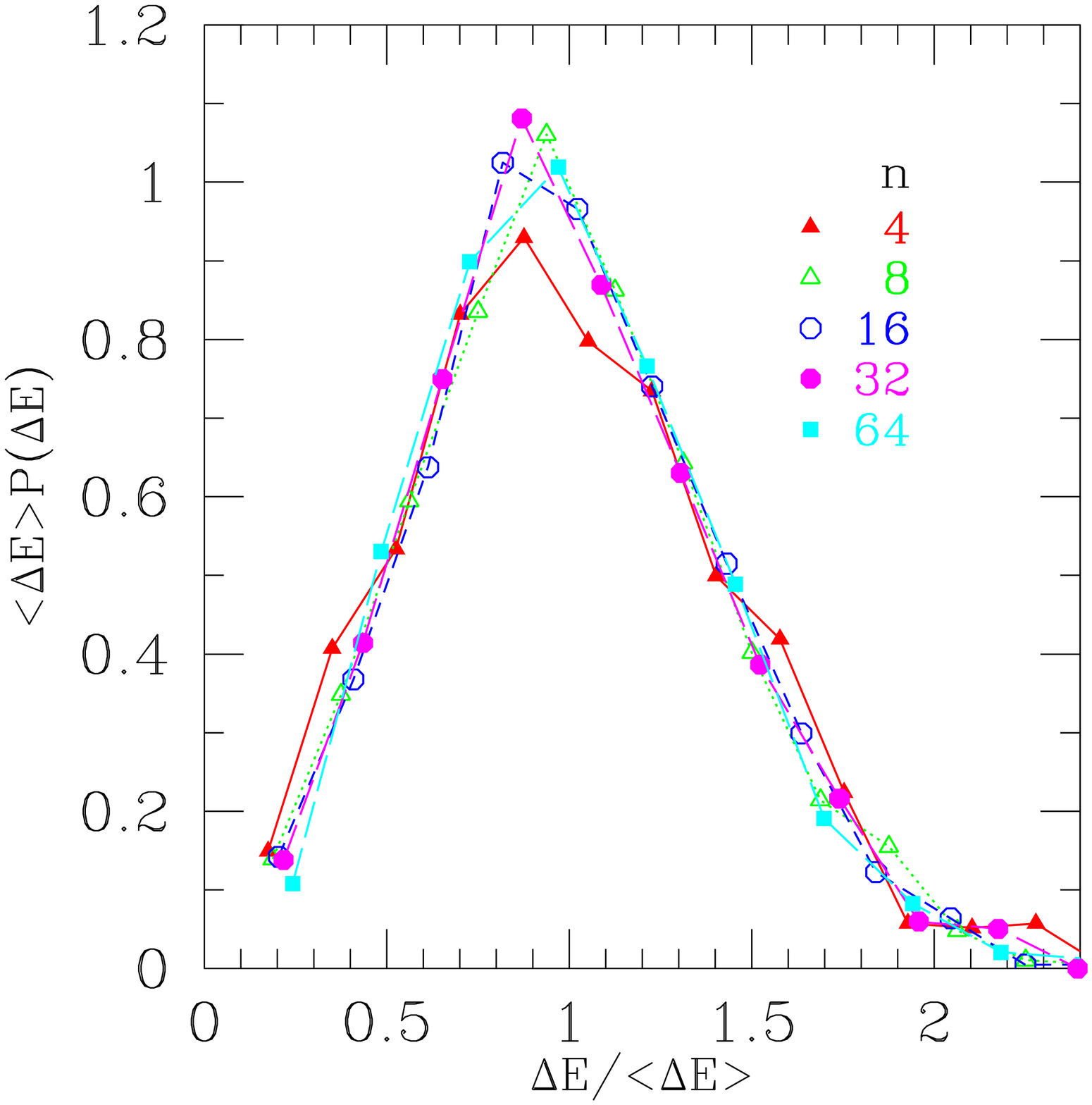}
\caption{The data in Fig.~\ref{dist} scaled by the mean 
energy $\langle \Delta E \rangle$, according to Eq.~(\ref{scaling}).}
\label{dist_scale}
\end{figure}

Fig.~\ref{dist} shows the distribution of the energy, $P(\Delta E)$,
of fixed size droplets for
different sizes with $L=64$. 
In Fig.~\ref{dist_scale}, we show 
that these distributions can be satisfactorily scaled using
the form
\begin{equation}
P(\Delta E) = \frac{1}{\langle \Delta E \rangle} 
{\cal P} \left( \frac{\Delta E}{\langle
\Delta E \rangle} \right),
\label{scaling}
\end{equation}
using the mean energy $\langle \Delta E \rangle$ which is plotted in
Fig.~\ref{fixed_vol}. The data indicates that the scaling function
${\cal P}(x)$ varies \textit{linearly} for small $x$. This result was
also obtained in three-dimensions\cite{lamarcq:02}, but 
differs from the usual assumption in droplet theory that the 
distribution has a finite weight at the origin.

\subsubsection{Fractal Dimension of the Droplets}
\label{sec:radii}
\begin{figure}
\includegraphics[width=\figurewidth]{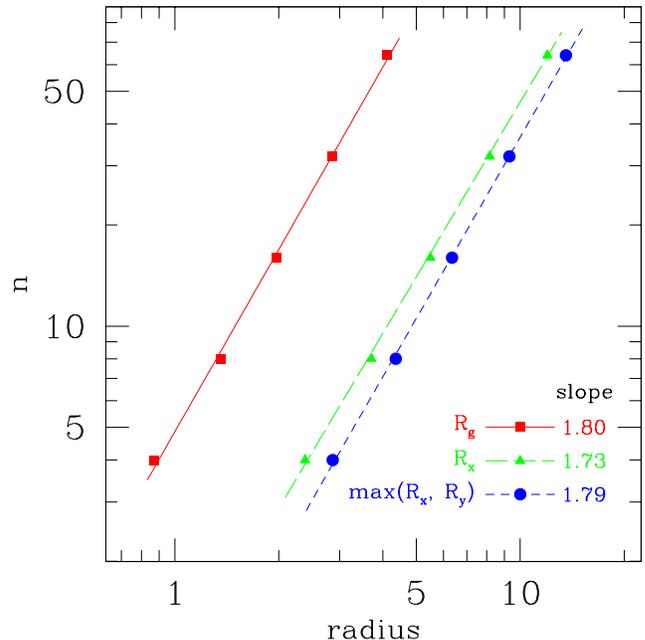}
\caption{
The number of spins in the droplet, $n$, as a function of various definitions
of the ``radius'', as described in the text. This is for droplets of a fixed
size. 
}
\label{radii}
\end{figure}

To determine the fractal dimension of the volume of the droplets through
Eq.~(\ref{d_f}), we need to
give a definition of their ``radius''.  We have done this in different ways
and checked that they give consistent results for $d_f$. We measured $R_x$ and
$R_y$, the maximum extent of the droplets in the $x$ and $y$
direction as well as max($R_x, R_y$). In addition we determined the radius of
gyration, $R_g$, of the droplet defined by:
\begin{equation}
R_g = \sqrt{{1\over n} \sum_{i \in {\rm cluster}}|\mathbf{r}_i -
\mathbf{r}_{\rm cm}|^2} ,
\end{equation}
where $\mathbf{r}_{\rm cm}$ is the position of
the center of mass of the cluster.

Figure~\ref{radii} shows the result of this analysis for $L=64$,
together with power law fits to the data. The agreement
between $R_g$ and ${\rm max}(R_x,R_y)$ is excellent. If we exclude the $L=4$
point, the fits give $d_f$ closer to 2. For example the fit to $R_g$ gives
$d_f \simeq 1.88$. Combining these values of $d_f$ with $\theta/d_f$ 
obtained from Fig.~\ref{fixed_vol} we get:
\begin{equation}
\theta \simeq  \left\{
\begin{array}{l}
-0.23 \qquad (\mbox{assuming } d_f = 2),  \\
-0.21 \qquad \mbox{(fit to } R_g),  \\
-0.22 \qquad \mbox{(fit to } R_g,\ \mbox{excluding } n=4). 
\end{array}
\right.
\end{equation}
We emphasize again that no trace of a more negative exponent
of the order of $-0.47$ can be detected in our data, 
thus in disagreement with the predictions by Moore\cite{moore:02} 
that this method should reveal the 
corrections to scaling included in Eq.~(\ref{deltaE}). 

\subsection{Clusters with a Range of Size}
\label{sec:scale_inv}
\begin{figure}
\includegraphics[width=\figurewidth]{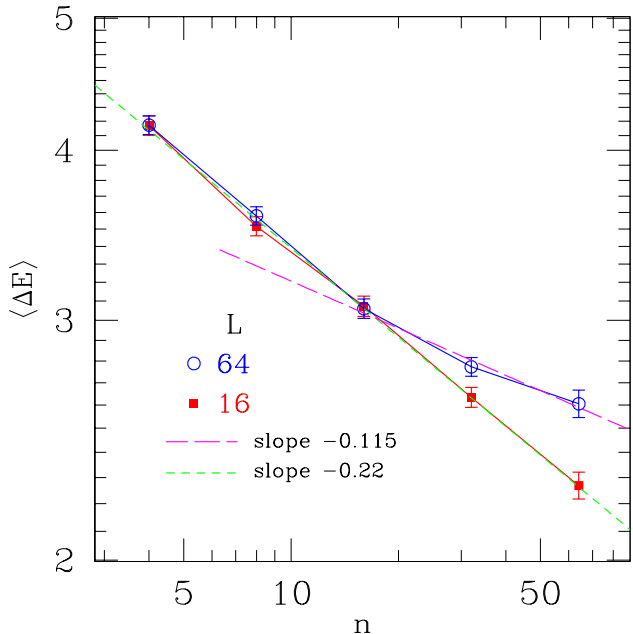}
\caption{
The mean energy of the scale-invariant droplets and system sizes
$L=16$ and $L=64$. For the size indicated as $n$,
the droplets are allowed to range in size between $n-1$ and $n/2$.
}
\label{scale_inv}
\end{figure}

In this section we allow the size of the droplets to vary over a factor of
two, more precisely between $n/2$ and $n-1$. This is
more
in the spirit of the droplet picture of Fisher
and Huse\cite{fisher:86,fisher:87,fisher:88}, where
droplets are defined, in $3d$, as objects with boundaries lying between 
a cube of linear size $\ell$ and another one of size $2\ell$.
We now perform the same analysis 
as above for these new clusters.

\subsubsection{Droplet Energies}
The mean droplet energy as a function of size $n$ is shown in
Fig.~\ref{scale_inv}. For the size specified as $n$, the droplets were allowed
to vary in size between $n-1$ and $n/2$. For very small sizes, where the data
for $L=16$ and 64 agree, the slope is about $-0.22$, considerably more
negative than the value of $-0.115$ found for droplets of fixed size in
Sec.~\ref{sec:fixed_vol}.  
For larger sizes,
where the data for $L=16$ differs from that for $L=64$, the slope of the
presumably more reliable $L=64$ data becomes less negative and is compatible
with $-0.115$, though there is not enough range of size to 
allow for a precise fitted value. 
If we insert $d_f = 2$, then, according to Eq.~(\ref{d_f}), this would
give a crossover from $\theta \simeq -0.44$ to $\theta \simeq -0.23$, not very
different from the crossover proposed in
Refs.~\onlinecite{moore:02,hartmann:02b}. 

However, it should be pointed out
that no such crossover was found in Sec.~\ref{sec:fixed_vol} for fixed size
droplets, and, furthermore, the values of the droplet radius $R$, discussed in
Sec.~\ref{sec:si_radii} are extremely small, of order one or two lattice
spacings, in the region where the data of Fig.~\ref{scale_inv}
suggests $\theta \simeq -0.44$, so that the meaning of a power law
correction of the type $L^{-\omega}$ for such sizes is not clear.
Note finally that the apparent fractal dimension for these very small clusters
is quite far from 2 (see below), so that, if we accept this value for $d_f$
the initial apparent 
stiffness exponent is around $-0.35$, rather far from 
the value $-0.47$ expected by Moore\cite{moore:02,hartmann:02b}.

Finally, we have to admit that we do not understand why the
crossover in Fig.~\ref{scale_inv} 
is absent when the size is exactly fixed, see Fig.~\ref{fixed_vol}.

\begin{figure}
\includegraphics[width=\figurewidth]{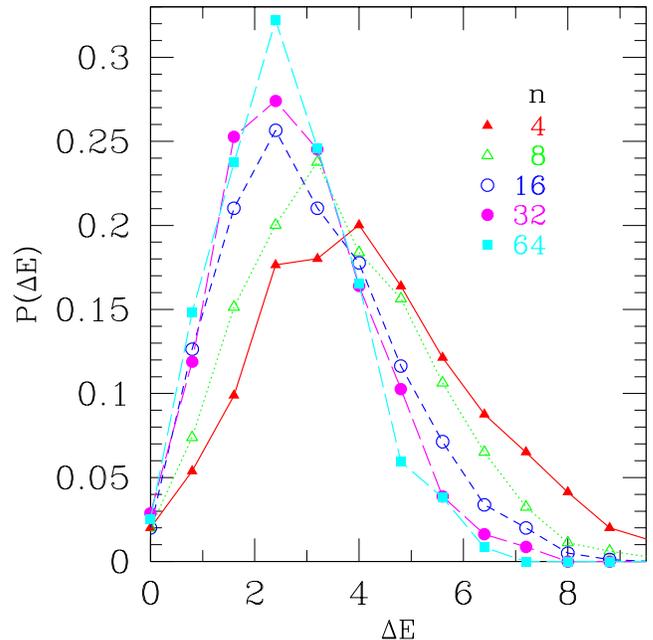}
\caption{
Distribution of the energy of
droplets of variable size, for different ranges of size for
$L=64$. For a size indicated as $n$, the droplets are allowed to
range in size between $n-1$ and $n/2$.
}
\label{dist_si}
\end{figure}

\begin{figure}
\includegraphics[width=\figurewidth]{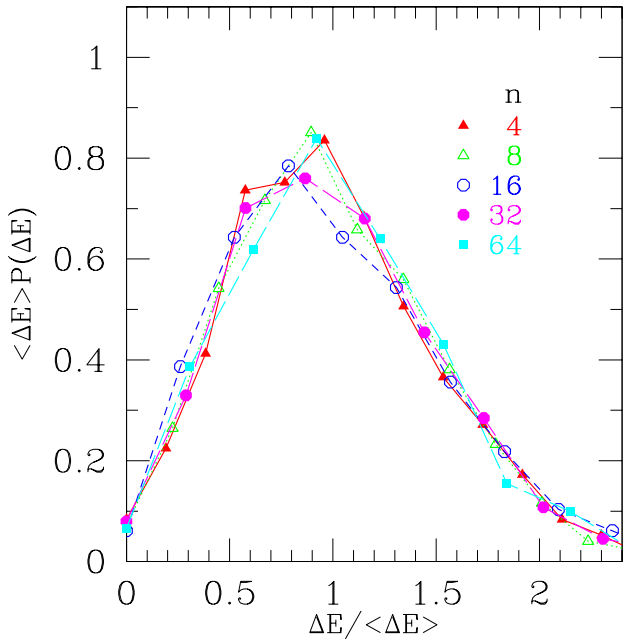}
\caption{
The data in Fig.~\ref{dist_si} scaled by the mean energy $\langle 
\Delta E \rangle$, according to Eq.~(\ref{scaling}).
}
\label{dist_si_scale}
\end{figure}

\subsubsection{Distribution of Droplet Energies}
The distribution of droplet energies is shown in Fig.~\ref{dist_si} and the
scaled data is plotted 
in Fig.~\ref{dist_si_scale}, again 
making use of Eq.~(\ref{scaling}). As for droplets of fixed size, the
distribution is essentially linear for small energy, though in this case there
is a very small, but definitely non-zero, intercept.

\subsubsection{Fractal Dimension of the Droplets}
\label{sec:si_radii}

Figure~\ref{radii_si} shows the various definitions of the radius of the
droplets, discussed in Sec.~\ref{sec:radii},
for different values of $n$. The data does not fit well a straight
line if the smallest size is included ($n=4$, corresponding to
droplets with 2 and 3 spins) so this has been omitted in the fits. The values
of $d_f$ for $R_g$ and max($R_x, R_y$) agree well, as for droplets of fixed
size, though the value of $d_f\ (\simeq 1.7)$ is a little lower than in that
case. However, the trend is for the slope to increase with increasing $n$,
this is particularly noticeable for the $R_g$ data, and so it is possible that
$d_f$ is actually equal to 2, as we concluded also in the case 
of the cluster of fixed size. 

\begin{figure}
\includegraphics[width=\figurewidth]{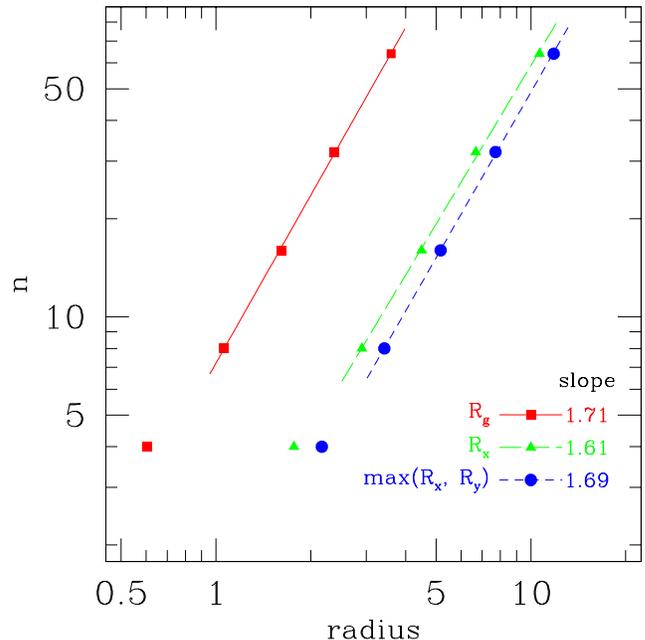}
\caption{
The number of spins in the droplet, $n$, as a function of various definitions
of the ``radius'', as described in the text. This is for droplets whose size
is allowed to vary between $n-1$ and $n/2$.
Data for $n=4$ has been excluded from the power law fit.}
\label{radii_si}
\end{figure}

\section{Conclusions}

The main motivation for our work is to
test Moore's proposal\cite{moore:02}
that contradictory results of the literature, both in 2 and
3 dimensions, for the determination of the stiffness
exponent $\theta$ in spin glasses were mainly due 
to ``corrections to scaling'' and captured by Eq.~(\ref{deltaE}). 
We have performed simulations similar to those of
Lamarcq et al.\cite{lamarcq:02,lamarcq:03} but in the 
$2d$ case, where this theoretical conjecture could be checked. 
From our results, we can draw the following conclusions.
\begin{enumerate}
\item Following exactly the procedure of Ref.~\onlinecite{lamarcq:02}, we do
not find any crossover in the behavior of the energy
of the cluster as a function of their size, up to $n=64$, 
contrary to Moore's prediction;

\item The value of the extracted stiffness exponent, $\theta \simeq
-0.23$,
is in fair agreement with domain wall calculations, $\theta = -0.28$. 
This result is clearly different from the $3d$ case, where
the droplet and domain wall exponents are found to be very different.

\item  For scale invariant clusters, a crossover not inconsistent with
(\ref{deltaE}) is obtained, but only for very small cluster sizes.

\end{enumerate}

As a conclusion, the problems we wanted to tackle in this paper unfortunately
remain unsolved. We still do not understand the $3d$ results by Lamarcq et
al.\cite{lamarcq:02} and we do not understand the origin of the two different
stiffness exponents reported in $2d$. Apart from very small sizes for one of
the two methods of generating clusters, we did not find evidence for the
crossover implied by Eq.~(\ref{deltaE}). The nature of corrections to scaling
in spin glasses therefore remains poorly understood.

Suppose we take the point of view that the differing claims obtained in
Refs.~\onlinecite{hartmann:02,hartmann:02b,picco:02,picco:03} and the present
work for the existence, or otherwise, of the proposed crossover in
two-dimensions, are evidence for a wide range of values of the amplitude of
the correction $B$ in Eq.~(\ref{deltaE}). It is then surprising that, to our
knowledge, \textit{all} estimates of $\theta$ from Monte Carlo simulations at
$T \ll T_c$ give $\theta$ (for droplets) close to (and consistent with) zero,
and different from $\theta$ for domain walls which is positive.
This is true for Ising spin glasses in three and four
dimensions\cite{katzgraber:01}, vector spin glass
models\cite{katzgraber:01a,katzgraber:01c}, and a one-dimensional model with
long range interactions\cite{katzgraber:03}.  In addition, various $T=0$
calculations \cite{krzakala:00,palassini:00,marinari:00} also find
$\theta\, \mbox{(droplets)}
\simeq 0$. If the amplitude of the correction term is highly non-universal, as
perhaps implied by the $2d$ results, it is surprising to us that the
parameters in Eq.~(\ref{deltaE}) systematically
conspire to give a droplet energy for, say, Monte Carlo simulations
which is independent of size over the range studied. Note too, that for the
$1d$ simulations\cite{katzgraber:03}, the range of sizes is very large, up to
$L=512$.

Of course, the usual criticism that the truth lies beyond the reach of
numerical simulations may be true. 
In this case, it is worth discussing the
relevance of physics which can only be seen on such astronomically long time
scales. From a \textit{mathematical} point of view, understanding
this physics is of fundamental importance. However, if the asymptotic behavior
would only be seen on time scales which are inaccessible to experiments as
well as to simulations, then the \textit{physical} relevance is rather 
limited.
In fact, experimental systems below the spin glass transition 
temperature $T_c$ are 
not truly in equilibrium but are only equilibrated up to some 
coherence length scale,
$\ell(t)$, which increases slowly, presumably logarithmically, 
with the time of the experiment $t$ and also depends on $T$.
One of us and Bouchaud\cite{berthier:02} have estimated that 
$\ell(T) \lesssim 10$ (in units of the spacing between spins) at reasonable 
experimental time scales at $T \simeq 0.5 T_c$.
Refs.~\onlinecite{jonsson:02,nordblad:03}
estimate a somewhat larger value (rather more than 20) 
at this
temperature and quite a lot larger\cite{jonsson:02}
closer to $T_c$. Hence a reasonable
estimate of $\ell(t)$ 
well below $T_c$ in experiments is around 10-20 times the spacing between
the spins, not very different from the sizes that can be simulated
numerically, and much less than the sizes which appear to be necessary to see
droplet behavior.

It is therefore possible that even if the droplet theory is
asymptotically correct, the region in which its predictions can be observed
quantitatively is not accessible experimentally.  Close to $T_c$ the length
scales which can be equilibrated in experiments may be much
larger\cite{berthier:02,jonsson:02}, 
but in this region critical fluctuations give
significant corrections to droplet behavior so \textit{even larger}
length scales are needed\cite{moore:98,bokil:00} to see droplet behavior
than at lower temperatures.

\begin{acknowledgments}
APY acknowledges support from the National Science Foundation under grant DMR
0086287 and the EPSRC under grant GR/R37869/01. He also thanks David
Sherrington for hospitality during his stay at Oxford when part of this work
was performed.
LB's work is supported 
by a European Marie Curie Fellowship No HPMF-CT-2002-01927, CNRS (France)
and Worcester College, Oxford. Numerical results were obtained 
on Oswell at the Oxford Supercomputing Center, Oxford University. 
\end{acknowledgments}

\bibliography{refs,comments}

\begin{thebibliography}{45}
\bibitem[{clu()}]{cluster_defn}
\bibinfo{note}{We use the terms ``cluster'' and ``droplet'' interchangeably in
  this paper}.
\end{thebibliography}

\end{document}